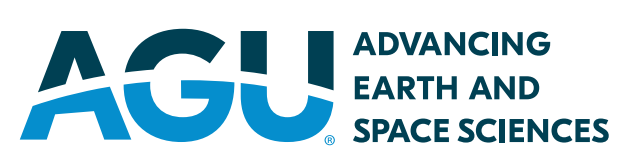

**Key Points:**

- We develop an attention-based simulator for multisite flood frequency, intensity, and duration informed by subdecadal climate evolution
- The model reproduces out-of-sample multivariate spatiotemporal dependencies for sequences of clustered extremes across multisite portfolios
- This framework addresses interannual-to-decadal flood risk for spatiotemporal insurance portfolio risk and financial planning

**Supporting Information:**

Supporting Information may be found in the online version of this article.

**Correspondence to:**

A. Nayak,
an3232@columbia.edu

**Citation:**

Nayak, A., Gentine, P., & Lall, U. (2026). Attention-based stochastic simulation of climate-informed spatiotemporal flood risk for multisite insurance portfolios. *Geophysical Research Letters*, *53*, e2026GL123701. https://doi.org/10.1029/2026GL123701

Received 17 APR 2026
Accepted 5 JUL 2026

**Author Contributions:**

**Conceptualization:** Adam Nayak, Pierre Gentine, Upmanu Lall
**Data curation:** Adam Nayak
**Formal analysis:** Adam Nayak
**Funding acquisition:** Adam Nayak
**Investigation:** Adam Nayak
**Methodology:** Adam Nayak
**Project administration:** Adam Nayak
**Resources:** Adam Nayak
**Software:** Adam Nayak
**Supervision:** Pierre Gentine, Upmanu Lall
**Validation:** Adam Nayak
**Visualization:** Adam Nayak
**Writing – original draft:** Adam Nayak

# Attention-Based Stochastic Simulation of Climate-Informed Spatiotemporal Flood Risk for Multisite Insurance Portfolios

**Adam Nayak[1,2,3], Pierre Gentine[1,2,3], and Upmanu Lall[1,2,4,5]**

[1]Department of Earth and Environmental Engineering, Columbia University, New York, NY, USA, [2]Columbia Water Center, Columbia Climate School, Columbia University, New York, NY, USA, [3]Learning the Earth with Artificial Intelligence and Physics (LEAP) National Science Foundation Center, Columbia University, New York, NY, USA, [4]School of Complex Adaptive Systems, Arizona State University, Tempe, AZ, USA, [5]The Water Institute, Arizona State University, Tempe, AZ, USA

**Abstract** Flood risk is correlated in space and time, challenging insurance systems that rely on diversification across assets. Financial instruments governing flood coverage are typically structured as 1–5-year contracts, exposing portfolios to climate-driven risk at interannual-to-decadal scales. Yet existing tools address climate risk either through seasonal forecasts extending only months or multidecadal projections misaligned with fiscal horizons, leaving a critical gap in actionable flood risk simulation. We introduce a multisite flood simulation framework combining attention-based analog retrieval with stochastic generation of multivariate flood frequency, intensity, and duration sequences. Applied to over 100 sites in the Mississippi River Basin, the model produces spatiotemporally coherent flood portfolios conditioned on interannual climate variability. Explainable AI attribution paired with wavelet analysis links simulated clustering to large-scale climate drivers, yielding physically interpretable flood clusters for portfolio-scale loss simulation. The framework provides plausible, out-of-sample flood risk catalogs for interannual-to-decadal insurance risk assessment and financial planning.

**Plain Language Summary** Floods often strike multiple places at once or in quick succession, straining insurance systems designed to spread risk across regions. Insurers largely price flood coverage around the fiscal year, yet available tools either forecast only months ahead or project risk decades into the future, leaving the interannual window that governs financial decisions unaddressed. Large-scale climate patterns like El Niño can synchronize floods across regions and years, but existing flood models do not often capture this variability at scales relevant to insurance pricing. We develop a simulation framework to fill this gap, combining machine learning and statistical methods to generate flood scenarios conditioned on interannual climate variability across more than 100 sites in the Mississippi River Basin. The model reproduces observed flood clustering in space and time while generating plausible interannual climate scenarios. Results are linked to physical climate drivers, making outputs understandable and directly useful for financial planning for flood risk.

## 1. Introduction

Floods are naturally correlated in space and time (Amonkar et al., 2023), challenging insurance systems that rely on risk diversification to buffer losses (Kousky & Cooke, 2012; Michel-Kerjan, 2010). Flood insurance in the United States has accumulated billions of dollars in debt (FEMA, 2024), which is largely attributable to space-time correlated losses induced by common hydrometeorological drivers (Nayak, Zhang, et al., 2025). Space-time clustered flooding exhibits fat-tailed (Bonnafous & Lall, 2021), nonstationary (Jain & Lall, 2001; Slater et al., 2021) risk distributions. Such risk is modulated by hydroclimatic regimes and moisture transport mechanisms that drive synoptic-scale flood clustering (Hirschboeck, 1988; Hwang et al., 2025; Nakamura et al., 2013). Clustered floods violate the independent and identically distributed (i.i.d.) generation assumptions underlying traditional flood frequency analysis used in actuarial pricing, which can lead to systematic risk underestimation (Nayak et al., 2024). Correlated risk is increasing under rising threats from flood co-occurrence (Kumar et al., 2025) and clustered regional flooding (Hwang and Lall, 2024; Nayak, Gentine, & Lall, 2025), and further compounded by a changing climate (Zscheischler et al., 2018) and population growth in floodplains (Tellman et al., 2021). Hence, accurately modeling these heavy-tailed, climate-driven risks remains an active challenge across hydrology, hydrometeorology, and hydroclimatology (Merz et al., 2022).

A fundamental gap in climate-conditional flood risk modeling today arises from a mismatch between the timescales of modeling climate variability and those most relevant for insurance decision-making. Insurance decision-making operates on much shorter timescales than traditional flood mitigation infrastructure design, with pricing decisions most commonly made at the fiscal year and reinsurance purchasing extending at most to 5-year contracts (Jaffree et al., 2008; Kunreuther & Michel-Kerjan, 2007). Robust precipitation signals from GCM projections do not emerge at subdecadal timescales (Hawkins and Sutton, 2011, 2012), and seasonal-to-subseasonal forecasts extend predictability only to months (Robertson and Vitart, 2018; Vitart et al., 2017). Climate-conditional approaches to flood risk that use downscaled GCM precipitation projections (Bates et al., 2023) tend to operate at longer decision timescales (multidecadal to centennial), missing the interannual-to-decadal window critical for shorter-term insurance decisions.

To address issues of spatiotemporally correlated flooding, multivariate extreme value theory has been widely applied to flood frequency analysis. Recent efforts to capture spatiotemporally correlated flooding include regression-based approaches (Diederen et al., 2019; Heffernan & Tawn, 2004; Keef et al., 2009; Quinn et al., 2019), copulas to simulate streamflow dependencies (Najibi et al., 2023; Yu et al., 2025), and climate-conditional approaches to multi-site flooding using downscaled GCM data (Bates et al., 2023). While these methods characterize the joint distribution of extremes, they typically rely on annual maxima formulations that neglect temporal sequences of intra-annual floods. Temporal sequences of floods are a common characteristic of synoptic-scale patterns of flooding driven by atmospheric blocking (Hirschboeck, 1988; Hwang et al., 2025; Nakamura et al., 2013). Temporally clustered extremes are increasing (Najibi and Devineni, 2018), and have been shown to result in increased economic impact due to recurrent and sequential impact (Bowers et al., 2024; Yang et al., 2026), jeopardizing flood insurance pooling due to lagged failures that occur within the same fiscal year (Nayak et al., 2024). Previous work has simulated point-based nonstationary temporal flood clustering (Nayak et al., 2024), and produced climate-conditional streamflow simulations at the interannual-to-decadal scale (Kwon et al., 2008, 2009; Lima et al., 2015; Najibi et al., 2023). However to our knowledge, no prior approaches (a) jointly capture interannual-to-decadal climate-conditioned variability, (b) represent temporally clustered sequences of flood frequency, intensity, and duration within years, and (c) generate multivariate, spatially coherent flood realizations for portfolio-scale risk assessment.

We introduce a new attention-based method of multivariate flood simulation to capture sequences of floods across a diverse portfolio of assets conditional on climate variability at the interannual-to-decadal timescale. Attention-based deep learning architectures have revolutionized sequence-to-sequence modeling by capturing long-range temporal dependencies (Vaswani et al., 2017). However, their heavy data requirements and tendencies to overfit and underrepresent out-of-sample extremes make heavy-tailed risk distributions at hydroclimatic timescales difficult to capture without long historical records (Sun et al., 2025). Nonparametric analog methods widely adopted in the field (Lall & Sharma, 1996) are well-suited to hydrologic simulation but can struggle with the curse of dimensionality at multisite scale (Amonkar et al., 2022). Drawing on advances in retrieval-augmented generation (RAG) (Lewis et al., 2020), which has popularized the use of learned embeddings to identify relevant analogs in data-limited settings (Zhang et al., 2025), we overcome these challenges to use a transformer encoder to compress high-dimensional hydroclimate variability into a low-dimensional climate state that enables coherent analog sequence retrieval at scale.

Our framework consists of three steps: (a) extraction of a multivariate hydroclimate signal, (b) ensemble forecasting of this signal at interannual-to-decadal scales, and (c) climate-conditional simulation of multivariate flood sequences. We demonstrate our model with a case study of multisite streamflow in the Mississippi River Basin. This framework enables the generation of nonstationary, climate-conditioned, spatiotemporally coherent flood sequences, introducing a new basis for portfolio-scale climate risk assessment in insurance and infrastructure systems.

## 2. Model

At a high level, our modeling goal is to simulate sequences of daily hydrologic extremes across multiple locations conditioned on low-frequency climate variability. We construct a regional low-frequency signal from monthly extremes, link it to large-scale climate indicators, and forecast its evolution using an attention–kNN approach. These forecasts then condition the generation of site-specific hydrologic statistics and flood sequences (Figure 1).

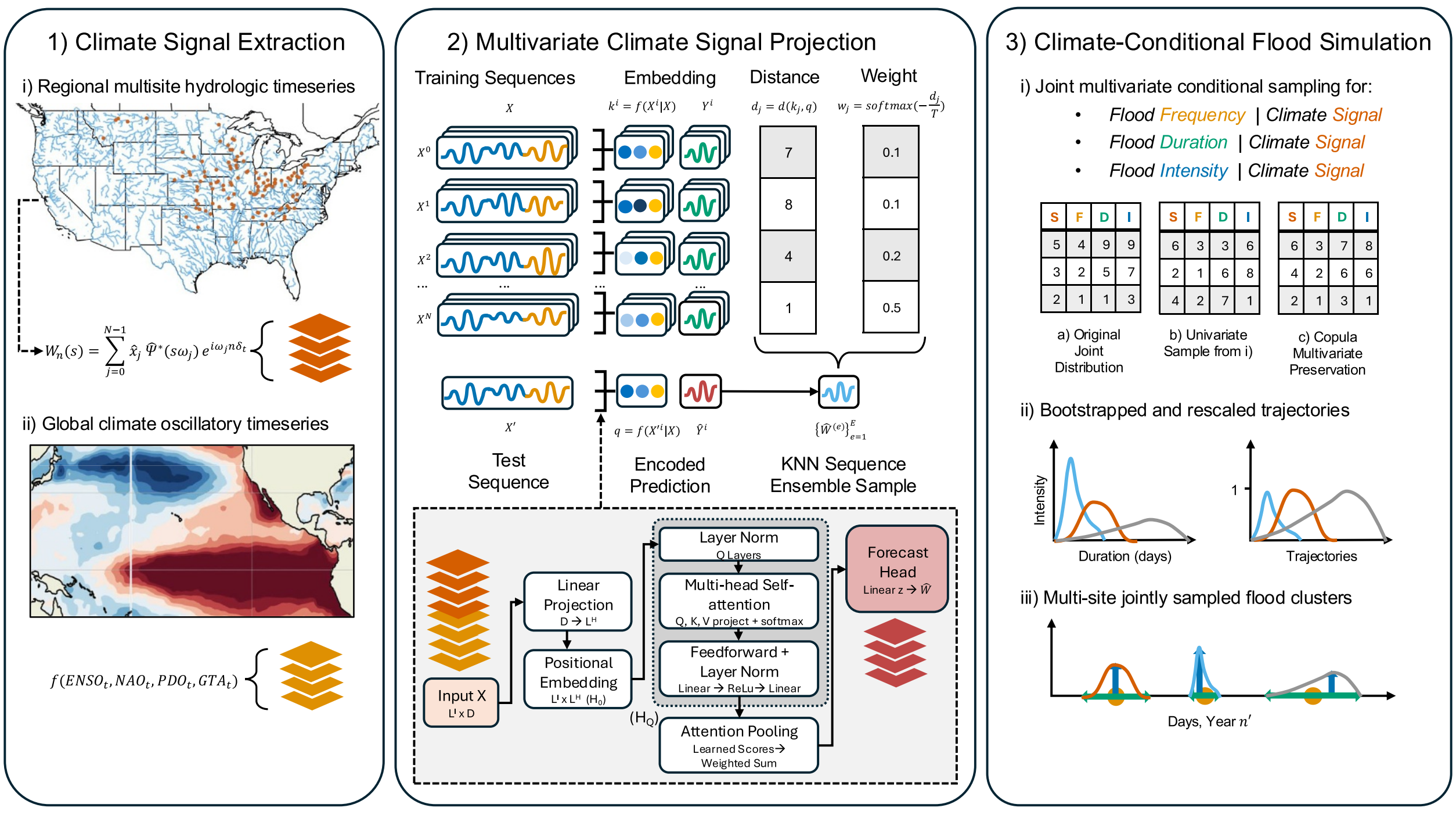

**Figure 1.** Framework for attention-based stochastic simulation of nonstationary, multisite extremes to evaluate climate-conditional clustered flood risk. Panels (2 and 3) adapted from (Nayak et al., 2024; Zhang et al., 2025).

The framework is modular, allowing alternative specifications at each stage depending on the application, while preserving a consistent structure linking climate variability to hydrologic extremes.

### 2.1. Climate Teleconnection Signal Extraction

We represent regional flood variability as arising from an underlying low-frequency hydroclimatic signal linking large-scale climate variability to flood risk across sites, similar to (Erkyihun et al., 2016; Kwon et al., 2007; Nowak et al., 2011; Rajagopalan et al., 2019). By examining the co-variability of regional hydrologic extremes with large-scale climate indicators, we aim to capture teleconnections between large-scale climate modes and local extremes. More formally, consider a $J$-dimensional multivariate daily hydrologic time series $x_{T,J}$ of length $T$ days and $J$ locations of interest. We compute the monthly maxima hydrologic time series $m_{N,J}$ over $N$ total months, then apply the continuous wavelet transform (CWT) to each univariate vector $j \epsilon \{1, \ldots, J\}$using the Morlet mother wavelet $\Psi_0$ (Torrence & Compo, 1998) (see Section 1 in Supporting Information S1 for full wavelet transform equations). Significant quasi-periodic bands are identified via red/white noise testing at the 0.95 significance level (detailed further in Supporting Information S1, Section 1.1), and reconstructed following previous studies to yield a multivariate low-frequency regional hydrologic signal $W_{N,J}$. We then select $C$ monthly climate indices representing global variability in pressure anomalies and sea surface temperatures that are spectrally coherent with the regional series, forming a climate teleconnection vector $O_{N,C}$. Climate indices $O_{N,C}$ can also be optionally denoised using wavelet spectral analysis, as done with the local monthly hydrologic time series $W_{N,J}$. The combined hydroclimatic signal is defined as $S_{N,D} = [W_{N,J}, O_{N,C}]$ such that $D = J + C$. In our case study, these include the Niño3.4 Index for the El Niño Southern Oscillation (ENSO), the North Atlantic Oscillation (NAO), the Pacific Decadal Oscillation (PDO), global temperature anomalies (GTA), and seasonality as considered in previous studies in the Mississippi River Basin (Muñoz & Dee, 2017; Olsen et al., 1999; Twine et al., 2005). This framework is modular: both the hydrologic representation $W_{N,J}$ and climate indicators $O_{N,C}$ can be specified in alternative ways at the discretion of the investigator, provided they capture low-frequency variability and teleconnections relevant to the region of interest.

### 2.2. Attention-Based Climate Signal Ensemble Forecasting

Next, we employ a hybrid deep learning framework to project our multivariate hydroclimatic signal forward. Deep learning has demonstrated impressive success in flood prediction (Kazadi et al., 2024; Kratzert et al., 2019; Nearing et al., 2022), yet many models tend to overfit, making heavy-tailed risk distributions difficult to capture without long historical records (Sun et al., 2025). Transformers are well-suited to this task through self-attention: a mechanism that learns which parts of a long input sequence are most relevant to each other, enabling efficient extraction of complex temporal dependencies across large multivariate data sets without assuming a fixed lag structure (Vaswani et al., 2017). We overcome the overfitting limitation by combining the transformer encoder with nonparametric analog retrieval conditioned on the climate state, drawing on retrieval-augmented generation (RAG) from natural language processing (Lewis et al., 2020). In this framework, a learned encoder maps inputs to an embedding space where nonparametric retrieval of observed neighbors informs the model output, without requiring the parametric component to memorize training examples. Prior work in multivariate time series forecasting introduced a transformer–kNN hybrid architecture in which encoder embeddings index a datastore of historical trajectories used to reweight forecast outputs (Zhang et al., 2025). We extend this framework by using encoder embeddings as a similarity metric for joint retrieval of historically observed multisite flood sequences, preserving spatial dependence structure by construction rather than imposing it parametrically. The approach harnesses transformers for pattern extraction across multivariate sequences while leveraging nonparametric statistics to preserve distributional fidelity under limited temporal observations.

First, we train an encoder to embed our multivariate climate signal into a meaningful lower dimensional space. Given an input window $X \epsilon R^{L_I x D}$ of the hydroclimatic signal $S_{N,D}$, the encoder maps the sequence through $Q$ multi-head self-attention layers and collapses it to a fixed-length climate state embedding $z \epsilon R^{L_H}$ via learned attention pooling (see Supporting Information S1, Section 1.2 for full encoder architecture detailing). The full hydroclimatic signal $S_{N,D}$ comprising wavelet-reconstructed hydrologic variability $W_{N,J}$ and climate teleconnection indices $O_{N,C}$ enters as input features, but only $W_{N,J}$ is used as the forecast target. The historic climate vector thus informs the embedding as a lagged covariate without being projected forward.

To ensure the embedding $z$ captures information relevant to future hydrologic signal evolution, the encoder is trained end-to-end with a linear forecast head that maps $z$ to a predicted signal trajectory over the forecast horizon $L_P$, optimized via mean squared error loss. In the application presented in this paper, we aim to capture interannual-to-decadal variability, and use an input sequence length $L_I$ of 60 months (5 years) for a variable forecast horizon $L_P$ of up to 96 months (i.e., the next 8 years). The forecast head serves only to shape the embedding during training and is not used at inference time.

Following training, a datastore $D = \{(z_r, y_r)\}_{r=1}^{R}$ is constructed from all training sequences, where $z_r$ is the encoder embedding of each input window and $y_r \epsilon R^{L_P \ x \ J}$ is the corresponding observed future trajectory of $W_{N,J}$. This indexes the full historical record in the learned embedding space, so that climate-state-similar historic periods can be retrieved efficiently at forecast time.

Lastly, we use an analog query of the embedded space to enumerate future signal trajectories. At forecast time the forecast head is set aside. The trained encoder embeds the current hydroclimate state into a query embedding $z_q$, which is used solely as a similarity metric to index $D$. The $K$ nearest neighbors are retrieved by Euclidean distance $d_r = ||z_q - z_r||_2$, with softmax sampling weights:

$$w_r = \frac{exp(-d_r/\tau)}{\sum_k^K exp(-d_k/\tau)}, \qquad r \, \epsilon \, \{1, \ldots, K\}$$

where $\tau$ controls retrieval sharpness (default $\tau = 1$). Each of $E$ ensemble members independently samples one analog trajectory $y_k$ from the $K$ nearest neighbors with probability $w_k$, yielding an ensemble $\{\hat{W}^{(e)}\}_{e=1}^{E}$ of plausible future signal trajectories.

Following model training, we employ integrated gradients for explainability (Sundararajan et al., 2017) across our fitted transformer encoder to derive the relative importance of different climate covariates within $O_{N,C}$ for the prediction of our future hydrologic signal sequences, providing a direct measure of large-scale climate influence on regional flood variability.

### 2.3. Climate-Conditional Stochastic Flood Cluster Generation

After reconstructing and forecasting the multivariate climate signal, we simulate clustered flood event sequences conditional on the ensemble climate signal forecast. Given an ensemble of forecast signal trajectories $\left\{\hat{W}^{(e)}\right\}_{e=1}^{E}$ from Section 2.2, we generate multisite flood sequences characterized by monthly frequency $\lambda$, peak intensity $\gamma$, and duration $\alpha$ at each site, conditioned on the projected hydroclimate state.

Flood events are identified by threshold exceedances in each daily hydrologic series $x_{T,j}$ for $j \epsilon \{1, \ldots, J\}$, with site-specific thresholds Thres $\epsilon R^{J}$ set to the bi-monthly peak exceedance value to capture the sequencing of seasonal peaks. For each historic month $n$ and site $j$, we extract frequency $\lambda_{n,j}$, max duration $\alpha_{n,j}$, and max peak intensity $\gamma_{n,j}$, of threshold exceedance sequences, forming a multisite reference repository linking flood characteristics to their corresponding signal values:

$$W'_{N,J} = \left\{W_{N,j}, \lambda_{N,j}, \alpha_{N,j}, \gamma_{N,j}\right\} \epsilon R^{NxJx4}$$

The joint analog datastore from Section 2.2 is augmented to include flood characteristics alongside signal trajectories:

$$D_{joint} = \left\{\left(z_r, \hat{y}_r\right)\right\}_{r=1}^{R}, \qquad \hat{y}_r \epsilon R^{L_P x(J+F)}$$

where $F = J\,x\,3$ reflects frequency, intensity, and duration at each of $J$ sites, and the embeddings $z_r$ are inherited directly from the trained encoder in Section 2.2. This nonparametric retrieval preserves the full observed spatial dependence structure of flood characteristics by construction. To enable generation of out-of-sample flood sequences not previously observed, we introduce new flood sequence samples using an empirical copula. Following the approach in (Nayak et al., 2024) and using the smoothed rank-based construction of (Lall et al., 2016; Najibi et al., 2023), marginal distributions of $\lambda_{N,j}, \alpha_{N,j}, \gamma_{N,j}$ and $W_{N,j}$ are fit independently per site with logspline density estimation, and their joint dependence is captured in a simulation pool of $N$ jointly distributed draws.

Next, we use the analog windows as probability-space conditioning targets. For each ensemble member $e$ and forecast month $n'$, the analog retrieval identifies a historically observed window whose embedding is nearest to the current climate state. The observed flood characteristic values $v*_{n',j}$ from that window are mapped through the fitted site marginal CDF to yield a target quantile:

$$u*_{n',j} = F(v*_{n',j})\ \epsilon\ [0, 1]$$

The copula simulation pool for site $j$ is pre-sorted by marginal CDF value. For each variable, the pool draw whose CDF position is nearest to $u*_{n',j}$ is selected via binary search to select the copula draw at the quantile level dictated by the climate analog, with optional Gaussian jitter $\varepsilon \sim N\left(0, \sigma^2\right)$ in probability space to introduce stochasticity across ensemble members sharing the same analog window (default $\sigma = 0.02$). For months where simulated frequency exceeds one, per-event intensity and duration are drawn from a separate multi-event copula pool using the same quantile matching procedure but reflecting historic exceedances that are not monthly maxima.

Lastly, simulated monthly event statistics are then disaggregated to daily flood trajectories via a kNN duration bootstrap of peak-intensity-normalized historical flood hydrographs (Nayak et al., 2024), detailed further in Section 6.2 in Supporting Information S1. We provide additional sampling options with a pure analog resampling structure and parametric copula fitting in Supporting Information S1, Section 1.3 and 1.4 including performance comparisons for all three sampling configurations.

## 3. Results

We evaluate the model using a case study of multisite streamflow across 117 USGS gauges in the Mississippi River Basin with a historical record of ≥90 years (see case study map in Supporting Information S1, Section 2). Climate covariates include four monthly indices sourced from NOAA with known teleconnections to regional flooding: Niño3.4 (ENSO), NAO, PDO, and global temperature anomalies (GTA) (Muñoz & Dee, 2017; Olsen et al., 1999; Twine et al., 2005). Performance is assessed using time-series cross-validation that preserves temporal dependence by training and testing on contiguous periods over the 1936–2025 record. Three cross-validation blocking schemes are employed: (a) continuous split-sample tests use consecutive 25-year training windows followed by 2-year forecasts, (b) block cross-validation uses progressively longer training periods of 50, 60, 70, and 80 years with 4, 5, 6, and 7 years forecast horizons, and (c) a standard time-series cross-validation window uses 81 years for training with an 8-year holdout. All experiments are performed jointly across all sites and include a 10% validation subset at the end of each training period. Overall, results indicate strong model performance across metrics of flood frequency, intensity, and duration. We structure this section to demonstrate model performance in (a) out-of-sample flood sequence ensemble calibration, (b) spatiotemporal portfolio dependence preservation, and (c) hydroclimatic driver attribution.

Simulated ensembles demonstrate out-of-sample performance, reproducing observed means and interquartile ranges across flood frequency, intensity, duration, and climate signal statistics in the held-out test periods (Figures 2a–2d). QQ envelopes confirm reliable distributional reproduction across sites, with median behavior centered on the 1:1 line for all variables (Figures 2e–2h). Inter-site spread widens in the upper tail as expected, reflecting increased quantile estimation uncertainty at extreme values with limited sample sizes, though median tracking remains close to 1:1 throughout. Climate signal shows the tightest overall band, consistent with residual interannual variability in the primary conditioning variable (panel h). Variance ratios are near unity across variables (Figures 2i–2l), with flood frequency exhibiting a slight positive bias of 3%–5%. Intensity and duration exhibit a slight negative bias of 6%–8% and 2%–3%. Inter-site variance is largest under the rolling split-sample scheme, consistent with shorter 25-year training windows providing less stable estimates of the full flood regime. Together these results indicate the framework generates well-calibrated flood ensembles that faithfully represent the historical distribution but perform beyond the observed record at the interannual-to-decadal timescale. We provide comparisons of performance with baseline models (univariate KNN, seasonal autoregressive, and parametric Neyman-Scott sampling) for further validation including QQ envelope and CRPS evaluations in Supporting Information S1, Section 3. Notably, the model outperforms baselines even when simulating jointly across CRPS and quantile replication (Figures S6 and S7 in Supporting Information S1).

Our model effectively preserves spatiotemporal dependence across the basin-wide gauge network. The simulated pairwise correlation matrix and tail dependency matrix reproduce the observed block structure across all 117 sites, with geographically proximate sub-basin clusters clearly preserved in both observed and simulated fields (Figures 3a–3d). Scatter plots of pairwise correlations confirm near 1:1 agreement between observed and simulated values across flood frequency, intensity, duration and climate signal, which are preserved in simulation (Figures 3e, 3f, 3i, and 3j). Tail dependency across pairwise correlations (Figures 3g, 3h, 3k, and 3l) is also retained, with some deterioration exhibited in flood frequency as expected being a sparse discrete variable (Figure 3h). Together these results demonstrate that the framework reproduces the pairwise spatial dependence structure of flooding across the portfolio, capturing the co-occurrence patterns that drive correlated loss accumulation (see example flood footprints in Supporting Information S1, Section 6.1). For further temporal evaluation, we provide lead-time dependency evaluation using CRPSS against a climatological baseline (Figures 3m–3o) as seen in (Araya et al., 2023). Interannual performance shows notable outperformance (consistently >0.2) of the climatological baseline without much deterioration out to 7 years in CRPSS across metrics of flood frequency, intensity, and duration. We provide seasonal assessment in Supporting Information S1, Section 4 (Figure S8 in Supporting Information S1).

Figure 4 demonstrates the model's ability to provide explainable insights into regional climate variability and teleconnections to hydrologic extremes. We classify gauges across the Mississippi into four main regions based upon USGS HUC codes: the Missouri River region (10), the Arkansas-White River region (11), the Upper Mississippi (07), and the Ohio and Tennessee River region (05/06). Wavelet signal processing combined with multivariate climate signal forecasting reveals significant hydrologic periodicities under red/white noise tests (Figures 4a–4h) and quantifies the relative importance of large-scale climate indices for predictability using integrated gradients (Sundararajan et al., 2017) (Figures 4i–4l). All gauges pick up significant sub-decadal periods of oscillation, often corresponding to heightened relative importance of ENSO to local forecasts across all basins. More specifically, the wavelet power spectra consistently pick up significant power at periods of 2–6 years and 8–12 years, consistent with the dominant periodicities of ENSO (2–6 years) and PDO/NAO (8–12 years) as supported by the integrated gradients attributions (Figures 4e–4h). The Missouri, Arkansas and White River basins exhibit a dominant periodicity across gauges of 11.6 years, while the Upper Mississippi is dominated by a periodicity of 8.6 years, and Ohio and Tennessee River basins show a dominant periodicity of 5.8 years, corresponding more strongly to ENSO cycles. The Arkansas-White-Red basins exhibit heightened attribution to PDO, while Ohio and Tennessee show stronger NAO attribution, suggesting a west-to-east shift in dominant Pacific versus Atlantic oceanic forcings across the Mississippi basin. In contrast, long-range variability explained by global temperature anomalies are weak across sites, which is reasonable considering the interannual-to-decadal period of evaluation. Insights highlight the model's capacity to link local hydrologic extremes to global climate drivers in an interpretable manner (see Supporting Information S1, Section 5 for PCA of the transformer embedding structure).

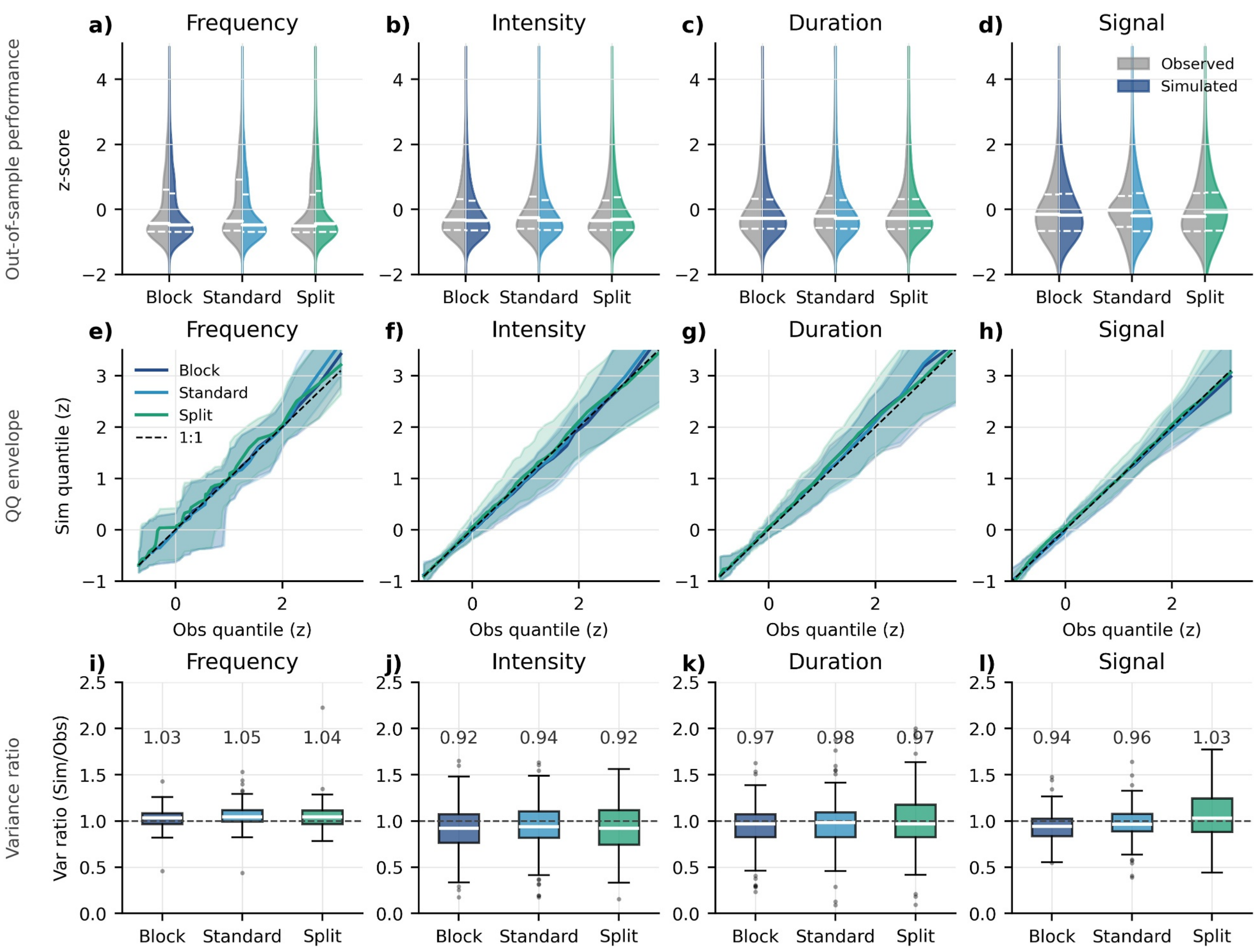

**Figure 2.** Out-of-sample test performance of simulated flood ensembles across 117 Mississippi River Basin gauges. (a–d) Split violin plots comparing z-scored distributions of flood frequency, intensity, duration, and climate signal between held-out observations (left) and simulations (right) across three cross-validation schemes. (e–h) QQ envelopes (10th–90th percentile across sites) of simulated versus observed quantiles for each variable, normalized per site to account for cross-basin magnitude differences. (i–l) Variance ratio (simulated/observed) across sites for each variable; values near unity indicate reproduction of observed variability, values above unity indicate ensemble over-dispersion.

## 4. Discussion

Evidenced by homeowners insurer market withdrawals (Eaglesham, 2023) and mounting national debt from flood disasters (FEMA, 2024; Nayak, Zhang, et al., 2025), large-scale spatiotemporally clustered damages (Nayak,

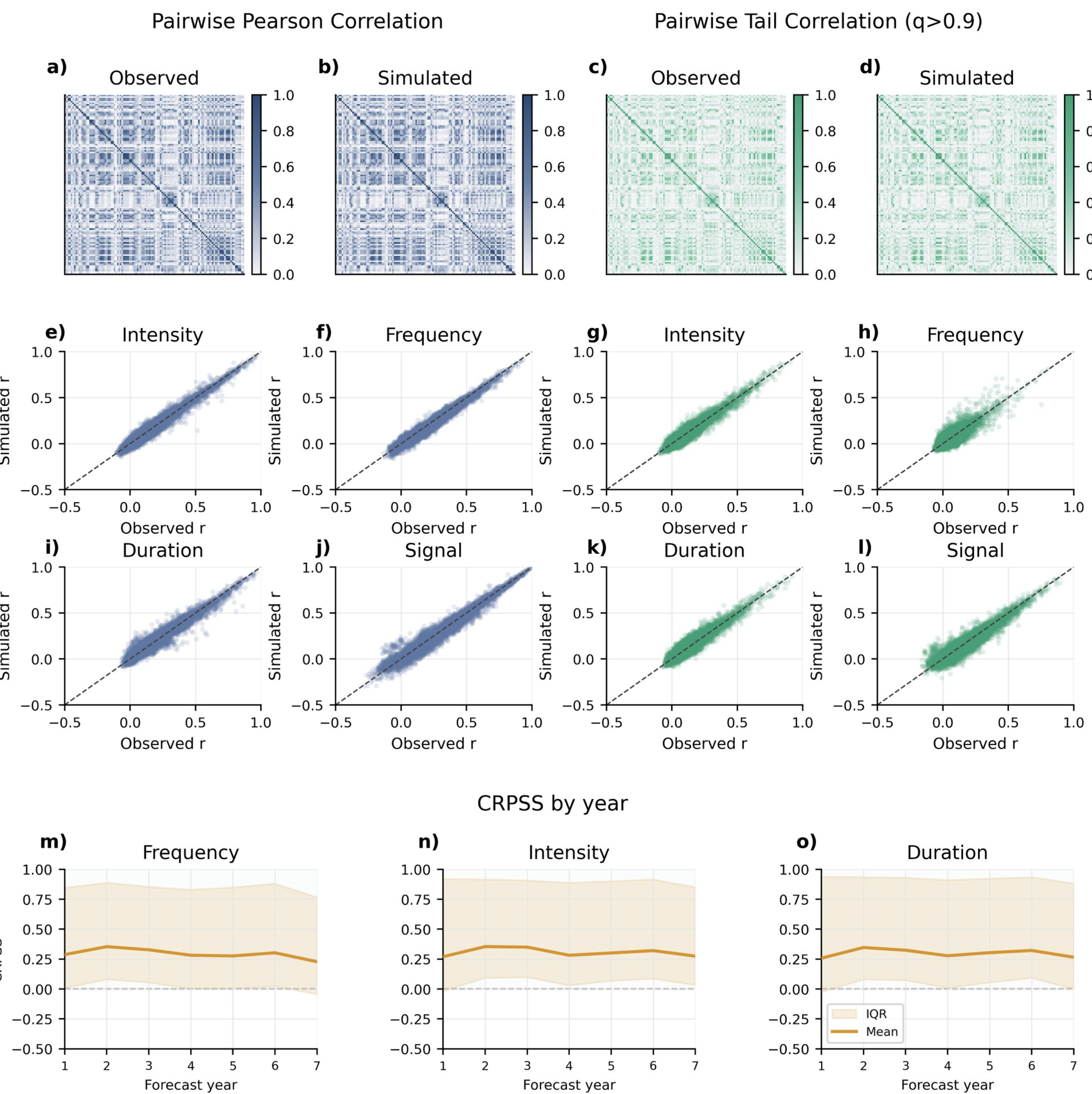

**Figure 3.** Reproduction of pairwise spatial dependence and tail dependence across 117 Mississippi River Basin gauges. Observed and simulated pairwise Pearson correlation matrices of the climate signal across sites, ordered by sub-basin (a–d). Scatter plots of observed versus simulated pairwise Pearson correlations for flood intensity, frequency, duration, and climate signal respectively (e–l); each point represents one gauge pair and the dashed line indicates 1:1 correspondence. CRPSS for interannual performance of flood frequency, intensity, and duration (m–o). Values above 0 indicate outperformance of the climatological baseline.

Zhang, et al., 2025) and correlated losses (Kousky & Cooke, 2012) jeopardize diverse insurance portfolios. In order to capture such risks, flood models must evolve to capture climate-driven spatiotemporal extremes at interannual-to-decadal timescales. We introduce an attention-based stochastic simulator for nonstationary, multisite extremes that evaluates climate-conditional clustered flood risk. The framework generates physically explainable, spatiotemporally coherent catalogs of flood sequences conditioned on interannual-to-decadal climate variability, directly supporting climate-conditional catastrophe risk assessment at portfolio scale. The framework is scalable with rapid simulation capacity, producing thousands of simulations in minutes across a CPU-driven train and test pipeline. Simulations provide easy integration with rapid flood inundation modeling such as LISFLOOD-FP (de Almeida and Bates, 2013; de Almeida et al., 2012), which takes distributed hydrographs as input that can be derived directly from our simulations (examples in Supporting Information S1, Section 7).

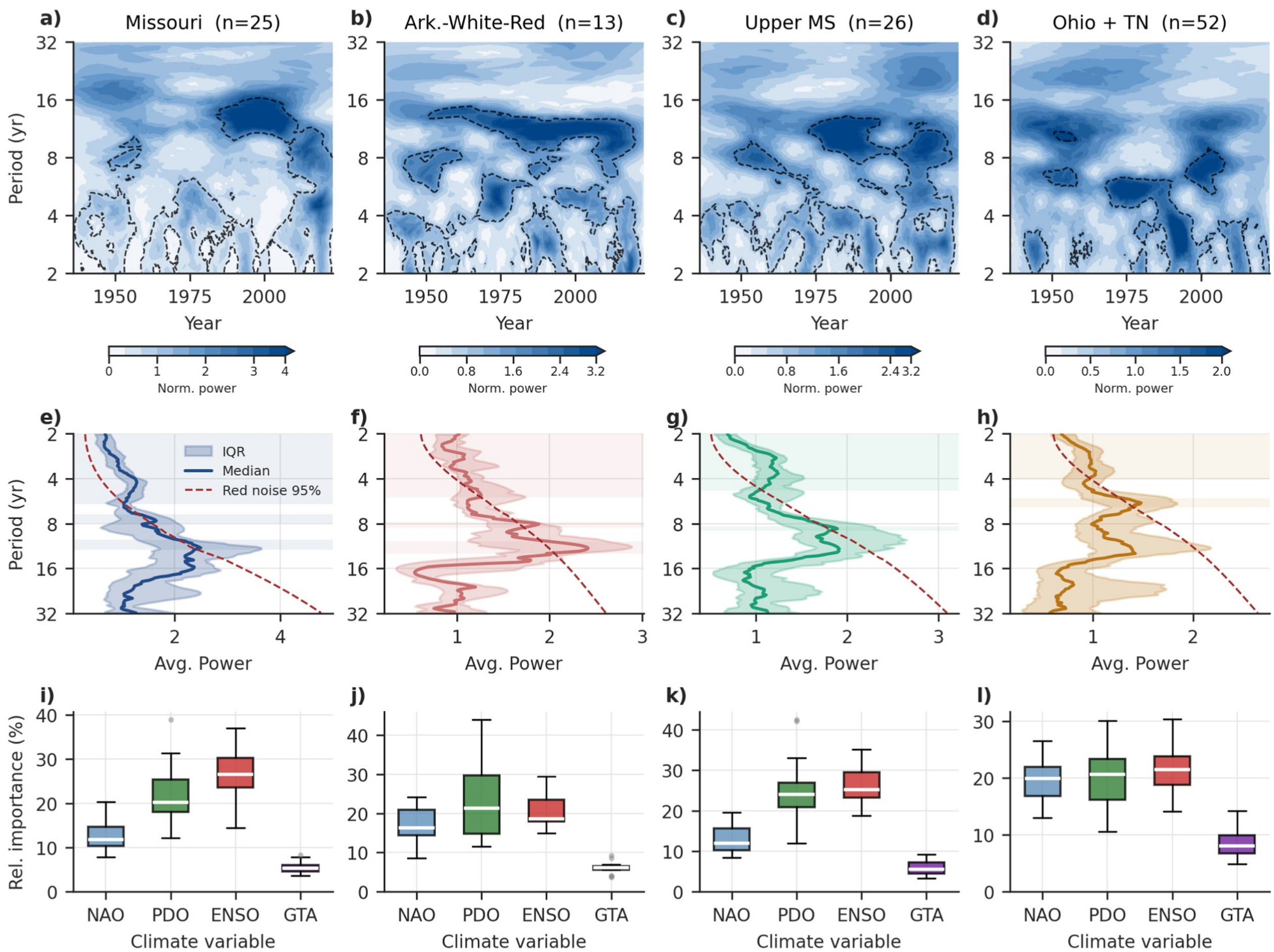

**Figure 4.** Explainable insights into climate drivers of flood generation for four Mississippi River subregions. Panels (a–d) show wavelet power spectra ($\lesssim$32 years) with significant periodicities outlined in the dashed regions; (e–h) display global spectra with significance from median red (dashed red) noise test with significant periodicities shaded; (i–l) present integrated gradient attributions relative importance (%) for ENSO, NAO, PDO, and GTA. Note ($n = 116$) as one singular gauge in the case study is in the lower Mississippi, hence it is omitted here.

Our Mississippi River Basin case study demonstrates the framework's capabilities, but the model can be enhanced under methodological extensions. For instance, alternative neural architectures could refine multivariate signal extraction, and future extensions may incorporate transfer learning with GCM ensemble projections of climate indices to further extend simulations beyond observed historic training data. Applications could include simulations of multisite precipitation extremes, or transfer learning to basins with limited observations. Future work can link moisture transport mechanisms and damage distributions to patterns of low-frequency variability, and consider adaptive pricing mechanisms to buffer correlated losses such as parametric tools (Khalil et al., 2007; Tellman et al., 2022). By accounting for climate-conditioned flood clustering in space and time, this framework advances interannual-to-decadal climate risk and catastrophe modeling, providing a practical tool for adaptive insurance design and financial risk management under climate variability at portfolio scale.

## Conflict of Interest

The authors declare no conflicts of interest relevant to this study.

## Availability Statement

All data used in this study is publicly available from the United States Geological Survey (USGS streamflow) at waterdata.usgs.gov and the National Oceanic and Atmospheric Association (NOAA climate indices) at psl.noaa.gov/data/climateindices/list. Open access code and data used in this study can be found at (Nayak, 2026).

**Acknowledgments**

Financial support for this research was provided by the National Science Foundation Graduate Research Fellowship Program (Grant DGE-2036197), and the Columbia Presidential Distinguished Fellowship from the Fu Foundation School of Engineering and Applied Sciences. Cloud computing resources were provided by the National Science Foundation's Science and Technology Center for Learning the Earth with Artificial Intelligence and Physics (LEAP) at Columbia University (Grant 2019625).